\documentclass[12pt,a4paper]{article}

\usepackage{epsfig}
\usepackage{amssymb}

\newcommand {\prd} [3] {Phys.~Rev.~D{\bf #1}, #2 (19#3)}

\newcommand {\asfer} [1] {\bar{{\tilde f}_{#1}} }
\newcommand {\aPn} [1] {a^{A^0}_{#1} }
\newcommand {\aSn} [1] {a^{H^0}_{#1} }
\newcommand {\ahn} [1] {a^{h^0}_{#1} }
\newcommand {\chin} [1] { {\tilde \chi}^0_{#1} }
\newcommand {\chip} [1] { {\tilde \chi}^+_{#1} }
\newcommand {\chipm} [1] { {\tilde \chi}^\pm_{#1} }
\newcommand {\fig} [1] {Fig.~\ref{#1}}
\newcommand {\gammu} {\gamma^\mu}
\newcommand {\gamnu} {\gamma^\nu}
\newcommand {\mchin} [1] { m_{{\tilde \chi}^0_{#1}} }
\newcommand {\mchip} [1] {m_{ {\tilde \chi}^+_{#1}} }
\newcommand {\msfer} [1] {m_{{\tilde f}_{#1}} }
\newcommand {\olz} [1] {O^L_{#1}{}'{}' }

\newcommand {\sfer} [1] {{\tilde f}_{#1} }
\newcommand {\tanb} {\tan \beta} 
\def\lsim{\raise0.3ex\hbox{$\;<$\kern-0.75em\raise-1.1ex\hbox{$\sim\;$}}}
\def\gsim{\raise0.3ex\hbox{$\;>$\kern-0.75em\raise-1.1ex\hbox{$\sim\;$}}}
\begin{document}

\title{
\begin{flushright} \small \rm
  hep-ph/9907377 \\ 
  UWThPh-1999-44 \\
  HEPHY-PUB 717/99 \\ \today
\end{flushright}
Large Higgs Boson Exchange Contribution in Three-Body Neutralino Decays}
\author{A.~Bartl$^1$, W.~Majerotto$^2$, and W.~Porod$^3$\\[1cm]
$^1$Institut f\"ur Theoretische Physik, Univ.~Wien, A-1090 Vienna, \\ Austria \\
$^{2}$ Institut f\"ur Hochenergiephysik der \"Osterreichischen \\
  Akademie der Wissenschaften, A-1050, Vienna, Austria \\
$^3$ Departament de F\'\i sica Te\`orica-IFIC, CSIC, Univ. de \\Val\`encia
Burjassot, Val\`encia 46100, Spain
}
\date{}
\maketitle

\begin{abstract}
We show that the Higgs boson exchange contribution can
be large in three-body decays of neutralinos even in the 
case of small $\tan \beta$. This enlarges the branching ratios for the
decays $\chin{2} \to \chin{1} \, b \, \bar{b}$ and 
$\chin{2} \to \chin{1} \, \tau^- \, \tau^+$. This is the case in the region
 of the parameter space where the two lightest
neutralinos are gaugino-like, the sfermions are heavier than 300~GeV,
and $m_{A^0} \lsim 200$~GeV.
\end{abstract}

\section{Introduction}

In the phenomenology of supersymmetric (SUSY) particles the neutralinos
$\chin{i}$ play a special r\^ole. In the Minimal Supersymmetric Standard
Model (MSSM) one has four neutralinos $\chin{i}$, $i=1,\dots,4$.
If R-parity is conserved $\chin{1}$
is considered to be the lightest stable SUSY particle (LSP). In $e^+ e^-$
or $p\, \bar{p}$ ($p\,p$) collisions neutralinos $\chin{i}$
as well as charginos $\chip{j}$, $j=1,2$, are 
produced directly or 
in cascade decays of heavier SUSY particles. Therefore,
a precise and complete theoretical knowledge of the decay properties
of neutralinos and charginos is mandatory for the understanding of all aspects
of SUSY phenomenology.

Neutralino decays have been extensively studied in the literature
\cite{Neutdecays}. In the following we reexamine the decays of $\chin{2}$. 
If two-body decays like $\chin{2} \to Z^0 \, \chin{1}$, 
$\chin{2} \to h^0 \, \chin{1}$, $\chin{2} \to f \, \asfer{}$ (with 
$\sfer{}$ being a slepton, sneutrino or squark) are kinematically not 
possible, the three-body decays 
\begin{eqnarray}
\label{eq:chi0tochi0}
\chin{i} &\to& \chin{j} \, f \, \bar{f} 
\end{eqnarray}
and
\begin{eqnarray}
\chin{i} &\to& \chip{j} \, f \, \bar{f}' 
\end{eqnarray}
will dominate. Here $f$ ($f'$) denote a Standard Model (SM) fermion.
 The decay (\ref{eq:chi0tochi0}) proceeds at tree level
via $Z^0$, $\sfer{}$ and Higgs boson exchange 
($h^0$, $H^0$, $A^0$ in the MSSM). 

Usually, the Higgs boson 
exchange is assumed to be small or even negligible, at least for small
or  moderate $\tanb$, where $\tanb = v_2/v_1$, $v_{1,2}$ being the
vacuum expectation values of the neutral Higgs fields in the MSSM. In this
note we will point out that Higgs exchange  can be important in a certain
domain of the SUSY parameters for 
$\chin{2} \to \chin{1} \, b \, \bar{b}$ and 
$\chin{2} \to \chin{1} \, \tau^- \, \tau^+$.
This is the case if $\chin{1}$ and $\chin{2}$ are gaugino-like ($M<|\mu|$)
and the sfermions are relatively heavy ($\msfer{} \gsim 300$~GeV). In this
case the $Z^0$ contribution will be suppressed because the $Z^0$ couples
only to the higgsino components of $\chin{1,2}$. The sfermion exchange
contribution will also be suppressed due to the heavy sfermion mass.
Then the Higgs boson exchange becomes important in the decays 
$\chin{2} \to \chin{1} \, b \, \bar{b}$ and 
$\chin{2} \to \chin{1} \, \tau^- \, \tau^+$ due to the corresponding
Yukawa couplings which are proportional to $m_b$ and $m_\tau$, respectively.
This enlarges the branching ratios for these decays, even for small $\tanb$,
and reduces the branching ratios for final states containing leptons and
quarks of the first and second generation. 
For large $\tanb$ this lepton non-universality
effect was already discussed in \cite{BaerLTb}.

Moreover, the Higgs exchange contribution depends on the parameters of the
Higgs sector. In the MSSM the $h^0$-$b$-$\bar{b}$ coupling depends on
the Higgs mixing angle $\sin \alpha$ \cite{GH}. 
We will show that Higgs boson exchange
plays a r\^ole at small $\tanb$ for $m_{A^0} \lsim 200$~GeV. For larger
values of $m_{A^0}$ $|\sin \alpha|$ becomes smaller and, therefore the 
Higgs boson exchange contribution becomes less important.
The Higgs boson exchange has, of course, implications for the neutralino 
search  at present and future colliders. For instance, the neutralino search at
LEP \cite{sugra1,WS} and the trilepton signal at TEVATRON due to 
$p\,\bar{p} \to \chipm{1} \chin{2} \to 3 \, l + p_{Tmiss}  
 \hspace{1mm}$ \cite{trilepton1,trilepton2} would be affected by the
Higgs boson contributions in the decays.


\section{Neutralino Decays}


In the following we are interested in the decays Eq.~(\ref{eq:chi0tochi0})
of $\chin{2}$ for the case
$|M| \ll |\mu|$. Then $\chin{2}$ decays mainly into 
$\chin{1} \, f \, \bar{f}$ because 
$m_{\chin{2}} \approx 2 m_{\chin{1}} \approx m_{\chip{1}} \approx M$.
The main contributions 
stem from the $Z$-boson and Higgs boson exchanges if the sfermions are 
relatively heavy ($m_{\tilde f} \gsim 300$~GeV).
For the following considerations it is instructive to have a look at the
amplitudes of these contributions:
\begin{eqnarray}
M_Z &=& - g^2 \, D^{\mu \nu}_Z \, \,\olz{ji} \,
   \bar{u}(p_l) \gammu (L_f P_L + R_f P_R) v(p_k) \, \, 
   \bar{u}(p_j) \gamnu  \gamma_5  u(p_i) \, , \nonumber \\ \\
M_{h^0} &=& g^2 \, D_{h^0} \, \, \ahn{ji} \, \ahn{f} \,
   \bar{u}(p_j) u(p_i) \, \,
   \bar{u}(p_l)  v(p_k) \, ,\\ 
M_{H^0} &=& g^2 \, D_{H^0} \, \, \aSn{ji} \, \aSn{f} \,
   \bar{u}(p_j) u(p_i) \, \,
   \bar{u}(p_l)  v(p_k) \, ,\\ 
M_{A^0} &=& - g^2 \, D_{A^0} \, \, \aPn{ji} \, \aPn{f} \,
   \bar{u}(p_j) \gamma_5 u(p_i) \, \,
   \bar{u}(p_l) \gamma_5 v(p_k) \, , 
\end{eqnarray}
where  $P_{L,R} = (1 \mp \gamma_5) / 2 $ and
$p_i, p_j, p_k, p_l$ are the momenta of $\chin{i}$, $\chin{j}$,
$\bar{f}$ and $f$, respectively. 
$D^{\mu \nu}_Z =
  [ - g^{\mu \nu} + (p_i-p_j)_\nu (p_i-p_j)_\mu / m^2_Z ]
 / [ (p_i-p_j)^2 - m^2_Z + i m_Z \Gamma_Z ] $ and
$D_{S^0} =
  1 / [ (p_i-p_j)^2 - m^2_{S^0} + i m_{S^0} \Gamma_{S^0} ]$ 
($S^0 = h^0, H^0, A^0$)
are the propagators.
The couplings are given by
\begin{eqnarray}
\olz{ij} &=& \frac{1}{2 \cos \theta_W}
 \left[ (N_{i4} N_{j4} - N_{i3} N_{j3}) \cos 2 \beta
      - (N_{i3} N_{j4} + N_{j3} N_{i4}) \sin 2 \beta \right]
\, , \label{coup:Znn} \nonumber \\ \\ 
L_f &=& I^3_f - e_f \sin^2 \theta_W \, , \\
R_f &=& - e_f \sin^2 \theta_W \, ,  \\ 
\ahn{ji} &=& q_{ji} \sin \alpha + s_{ji} \cos \alpha \, , \\
\aSn{ji} &=& - q_{ji} \cos \alpha + s_{ji} \sin \alpha \, , \\
\aPn{ji} &=& - q_{ji} \sin \beta + s_{ji} \cos \beta \, , \\
q_{ji} &=&  \frac{1}{2 \cos \theta_W}
   \left[ ( N_{i3} \cos \beta + N_{i4} \sin \beta ) N_{j2}
        + ( N_{j3} \cos \beta + N_{j4} \sin \beta ) N_{i2} \right] \, ,
\nonumber \\ \\
s_{ji} &=&  \frac{1}{2 \cos \theta_W}
   \left[ ( N_{i4} \cos \beta - N_{i3} \sin \beta ) N_{j2}
        + ( N_{j4} \cos \beta - N_{j3} \sin \beta ) N_{i2} \right] \, ,
\nonumber \\  \\
\ahn{f} &=& \frac{m_f \sin \alpha}{2 \, m_W \cos \beta} \, , \hspace{1cm} 
    \aSn{f} = \frac{-m_f \cos \alpha}{2 \, m_W \cos \beta} \, , \hspace{1cm}
   (f=b,\tau) \, ,  \label{coup:Hbb} \\
\aPn{f} &=& \frac{m_f \tan \beta}{2\, m_W} \,\, \, (f=b,\tau) \, 
\label{coup:Abb},
\end{eqnarray}
where $I^3_f$ is the 3rd component of the isospin, $e_f$ is the charge of the
fermion in units of the positron,
and $N_{ij}$ is the neutralino mixing matrix in the 
notation of \cite{Bartl89a},

 The important point here is that the $Z$-boson only couples to the
higgsino components of the neutralinos, Eq.~(\ref{coup:Znn}).
Therefore, for $|M| \ll |\mu|$ the $Z$-$\chin{2}$-$\chin{1}$ coupling
can be rather
small and the Higgs boson exchange contribution can become important 
due to the $h^0$-$b$-$\bar{b}$ coupling. This implies
that the products of $Z$-couplings $L_b \, \olz{21}$ and $R_b \, \olz{21}$ 
can be smaller than the product of Higgs couplings
$\ahn{ji} \, \ahn{b}$ as is demonstrated in the next section.

\section{Numerical Results}

In  our numerical analysis we have taken $\sin^2 \theta_W = 0.2315$, 
$\alpha(m_Z) = 1/128$, 
$m_Z =91.187$~GeV, $m_\tau = 1.777$~GeV, $m_b = 5$~GeV, and $m_t = 175$~GeV. 
In the $b$ Yukawa couplings, Eqs.~(\ref{coup:Hbb}) and (\ref{coup:Abb}), 
we have taken the
running $m_b$ mass according to the renormalization group equations
\cite{Drees90b}.
Moreover, we have fixed
$M_{\tilde F} = 500$~GeV ($F=E,L,D,Q,U$), $A_i = 100$~GeV ($i=\tau,b,t$), 
$M=120$~GeV, and $M'= 5/3 \tan^2 \theta_W M$. With this choice of 
parameters the sfermion exchange is  suppressed relative to 
the $h^0$ exchange. 
For the radiative corrections to the $h^0$ and $H^0$ masses 
and their mixing angle $\alpha$ we use the formulae of 
Ref.~\cite{ellis}; 
for those to $m_{H^+}$ we follow Ref.~\cite{brignole}.\footnote{
Notice that \cite{ellis,brignole} have a sign convention 
for the parameter $\mu$ opposite to the one used here.}

\begin{figure}[ht]
\setlength{\unitlength}{1mm}
\begin{picture}(150,115)
\put(-1,-6){\mbox{\epsfig{figure=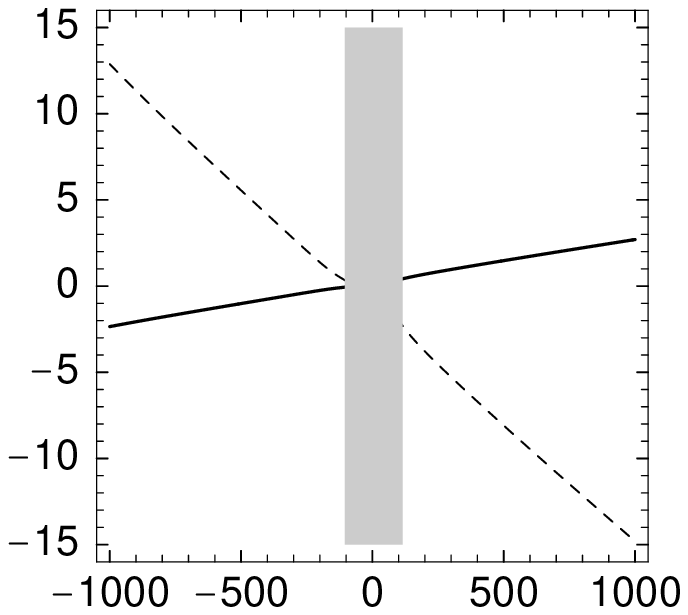,
                          height=11.cm,width=13.5cm}}}
\put(133,-4){\makebox(0,0)[br]{{\large $\bf \mu$}~[GeV]}}
\end{picture}
\caption[]{Ratios of couplings 
$(\ahn{21} \, \ahn{b}) / (L_b \, \olz{21})$ (full line) and
$(\ahn{21} \, \ahn{b}) /$ $(R_b \, \olz{21})$ (dashed line)
as a function of $\mu$ for $\tan \beta = 4$ and $m_{A^0} = 125$~GeV.
The other parameters are given in the text and the couplings are defined in
Eqs.~(\ref{coup:Znn}) - (\ref{coup:Hbb}). The grey area will
be covered by LEP2 ($\mchip{1} \le 95$~GeV).
}
\label{fig:CoupMu}
\end{figure}

\fig{fig:CoupMu}
shows the ratios of couplings 
$(\ahn{21} \, \ahn{b}) / (L_b \, \olz{21})$ (full line) and
 $(\ahn{21} \, \ahn{b}) / (R_b \, \olz{21})$ (dashed line)
as a function of $\mu$ for $\tan \beta = 4$ and $m_{A^0} = 125$~GeV. In the
regions with gaugino-like $\chin{1,2}$ ($|\mu|\ge M$) these ratios are always
larger than 1 and the absolute value of 
$(\ahn{21} \, \ahn{b}) / (R_b \, \olz{21})$ goes up to 15.

\begin{figure}[ht]
\setlength{\unitlength}{1mm}
\begin{picture}(150,115)
\put(-1,-4){\mbox{\epsfig{figure=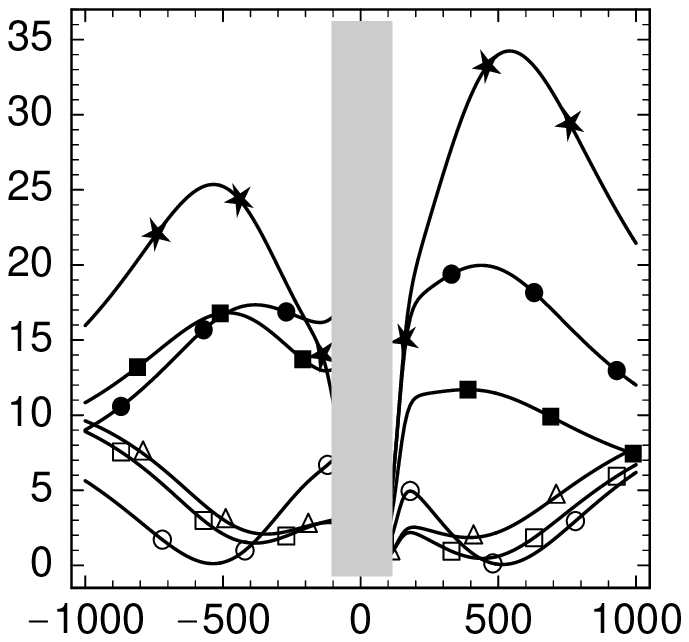,
                          height=11.cm,width=13.5cm}}}
\put(0,104){\makebox(0,0)[bl]{{\large
             $\bf BR$~[\%]}}}
\put(133,-4){\makebox(0,0)[br]{{\large $\bf \mu$}~[GeV]}}
\end{picture}
\caption[]{Branching ratios of $\chin{2}$ as a function of $\mu$ for 
$\tan \beta = 4$ and $m_{A^0} = 125$~GeV.
The other parameters are given in the text.
The graphs correspond to following transitions:
$\circ \, \chin{2} \to \chin{1} \, \nu_l \, \nu_l$ ($l=e,\mu$, or $\tau$), 
$\square \, \chin{2} \to \chin{1} \, l^+ \, l^-$ ($l=e$ or $\mu$), 
$\triangle \, \chin{2} \to \chin{1} \, \tau^+ \, \tau^-$, 
$\blacksquare \, \chin{2} \to \chin{1} \, q \, \bar{q}$ ($q=u$ or $c$), 
$\bullet \,  \chin{2} \to \chin{1} \, q \, \bar{q}$ ($q=d$ or $s$), and 
$\star \,  \chin{2} \to \chin{1} \, b \, \bar{b}$. The grey area will
be covered by LEP2 ($\mchip{1} \le 95$~GeV).
}
\label{fig:BrMu}
\end{figure}

In \fig{fig:BrMu} we show the branching ratios of 
$\chin{2} \to \chin{1} \, f \, \bar{f}$ as a function of $\mu$  where $f$
denotes $\nu_l$, $e$, $\tau$, $u$, $d$, or $b$. 
The other parameters are as in \fig{fig:CoupMu}.
The final state with $b \, \bar{b}$ clearly dominates for $|\mu| > M$
due to the dominance of the Higgs exchange. Also the 
$\chin{1} \, \tau^- \, \tau^+$ final state is enhanced compared to the
$\chin{1} \, l^- \, l^+$ ($l=e,\mu$) final states for the same reason.
The branching ratio for $\chin{2} \to \chin{1} \, b \, \bar{b}$ has maxima
near $\mu = \pm 500$~GeV. This can be explained by the fact that the gaugino
components of $\chin{1,2}$ are quite generally rising with $|\mu|$ and are
constant for $|\mu|\gsim 500$~GeV, whereas the higgsino components are
decreasing with $|\mu|$.

\begin{figure}[ht]
\setlength{\unitlength}{1mm}
\begin{picture}(150,115)
\put(-1,-5){\mbox{\epsfig{figure=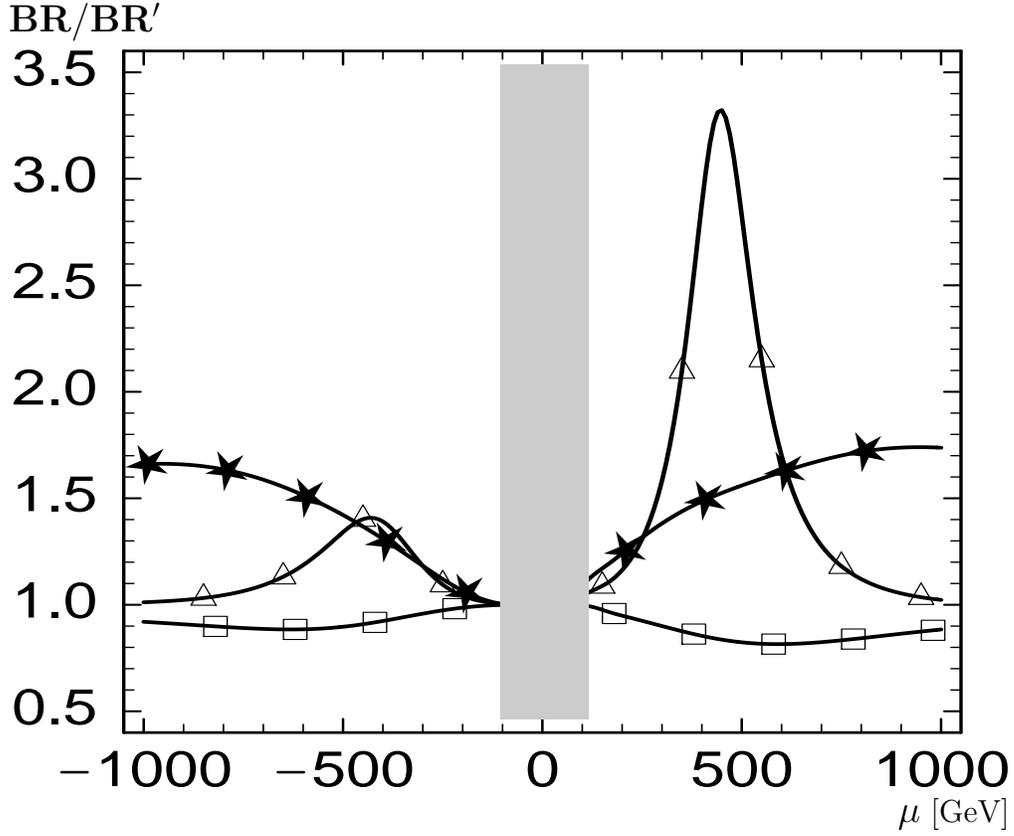,
                          height=11.cm,width=13.5cm}}}
\put(0,101){\makebox(0,0)[bl]{{\large
             $\bf BR/BR'$}}}
\put(133,-4){\makebox(0,0)[br]{{\large $\bf \mu$}~[GeV]}}
\end{picture}
\caption[]{
Ratio $BR(\chin{2} \to \chin{1} \,  f \, \bar{f}) /
        BR'(\chin{2} \to \chin{1} \, f \, \bar{f})$
 as a function of $\mu$ for 
$\tan \beta = 4$ and $m_{A^0} = 125$~GeV.
The other parameters are given in the text. 
In case of $BR$ ($BR'$) we have included 
(neglected) the Higgs boson
contributions to the partial decay widths.
The graphs correspond to:
$\square \, e$ or $\mu$,  
$\triangle \, \tau$, and
$\star \, b$. 
}
\label{fig:RatioMu}
\end{figure}

 \fig{fig:RatioMu} shows the ratio 
$BR(\chin{2} \to \chin{1} \, f \, \bar{f}) /
        BR'(\chin{2} \to \chin{1} \, f \, \bar{f})$ [$f=e,\mu,\tau,b$]
 as a function of $\mu$, where in $BR$ ($BR'$) we have included 
(neglected) the Higgs boson
contributions to the partial decay widths. The $e$ and $\mu$ channels can be 
reduced by up to 20\%. The branching ratios for the $\tau$ and $b$ channels
are enhanced. The maxima for the $\tau$ channel near $\mu = \pm 500$~GeV
can be explained in the following way: 
$BR'(\chin{2} \to \chin{1} \, \tau^+ \, \tau^-)$ has minima at these
$\mu$ values due to a 
destructive
$Z$-${\tilde{\tau}_i}$ interference, and 
$BR(\chin{2} \to \chin{1} \, \tau^+ \, \tau^-)$ 
is enhanced due to Higgs exchange.

\begin{figure}[ht]
\setlength{\unitlength}{1mm}
\begin{picture}(150,115)
\put(-1,-2){\mbox{\epsfig{figure=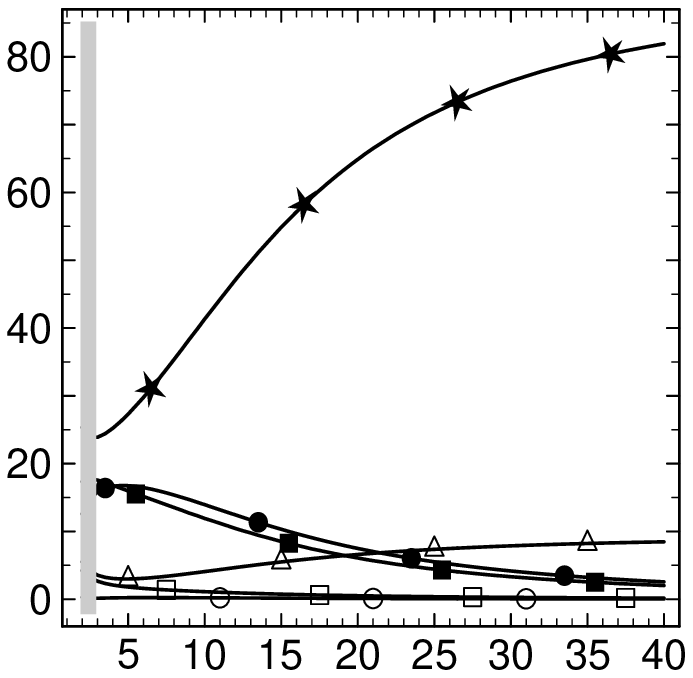,
                          height=11.cm,width=13.5cm}}}
\put(0,109){\makebox(0,0)[bl]{{\large
             $\bf BR$~[\%]}}}
\put(133,-4){\makebox(0,0)[br]{{\large $\bf \tanb$}}}
\end{picture}
\caption[]{Branching ratios of $\chin{2}$ 
as a function of $\tan \beta$ for
 $\mu = -500$~GeV and $m_{A^0} = 125$~GeV.
The other parameters are given in the text.
The graphs correspond to following transitions:
$\circ \, \chin{2} \to \chin{1} \, \nu_l \, \nu_l$ ($l=e,\mu$, or $\tau$), 
$\square \, \chin{2} \to \chin{1} \, l^+ \, l^-$ ($l=e$ or $\mu$), 
$\triangle \, \chin{2} \to \chin{1} \, \tau^+ \, \tau^-$, 
$\blacksquare \, \chin{2} \to \chin{1} \, q \, \bar{q}$ ($q=u$ or $c$), 
$\bullet \,  \chin{2} \to \chin{1} \, q \, \bar{q}$ ($q=d$ or $s$), and 
$\star \,  \chin{2} \to \chin{1} \, b \, \bar{b}$. In the gray area is
$m_{h^0} < 90$~GeV.}
\label{fig:BrTbn}
\end{figure}

\begin{figure}[ht]
\setlength{\unitlength}{1mm}
\begin{picture}(150,115)
\put(-1,-2){\mbox{\epsfig{figure=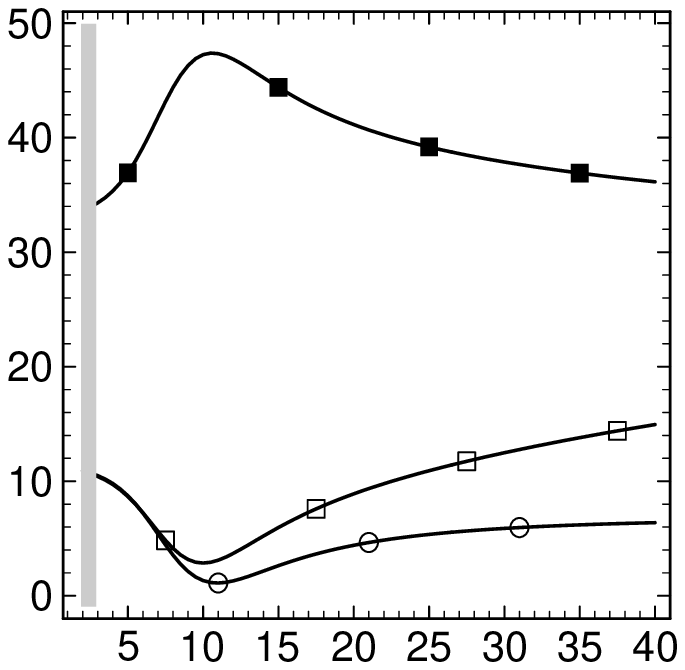,
                          height=11.2cm,width=13.5cm}}}
\put(0,109){\makebox(0,0)[bl]{{\large
             $\bf BR$~[\%]}}}
\put(133,-4){\makebox(0,0)[br]{{\large $\bf \tanb$}}}
\end{picture}
\caption[]{Branching ratios of $\chip{1}$
as a function of $\tan \beta$ for
 $\mu = -500$~GeV and $m_{A^0} = 125$~GeV.
The other parameters are given in the text.
The graphs correspond to following transitions:
$\circ \, \chip{1} \to \chin{1} \, l^+  \, \nu_l$ ($l=e$ or $\mu$), 
$\square \, \chip{1} \to \chin{1} \, \tau^+  \, \nu_\tau$, and
$\blacksquare \, \chip{2} \to \chin{1} \, q' \, \bar{q}$
 ($(q',q) = (u,d)$ or $(c,s)$).
 In the gray area is
$m_{h^0} < 90$~GeV.}
\label{fig:BrChipTbn}
\end{figure}

\fig{fig:BrTbn} shows the $\tanb$ dependence of 
$BR(\chin{2} \to \chin{1} \, f \, \bar{f})$ for $\mu = -500$~GeV. The 
rising of the $b \bar{b}$ and $\tau^+ \tau^-$ channels is mainly due to the
increase of the $b$- and $\tau$- Yukawa couplings. We also show in
\fig{fig:BrChipTbn} the $\tanb$ dependence of the branching ratios for the 
$\chip{1}$ decays. Note that 
$BR(\chip{1} \to \chin{1} \, \tau^+ \, \nu_\tau)$ is nearly twice 
$BR(\chip{1} \to \chin{1} \, e^+ \, \nu_e)$ for large $\tanb$.
The maximum (minima) of the 
$BR(\chip{1} \to \chin{1} \, q \, \bar{q}')$ 
($BR(\chip{1} \to \chin{1} \, e^+ \, \nu_e)$ 
and $BR(\chip{1} \to \chin{1} \, \tau^+ \, \nu_\tau)$) 
at $\tanb\sim 10$ is due
to a positive $W$-$\tilde{q}$ (negative $W$-$\tilde{l}$ and 
$W$-$\tilde{\tau}$) interference.

\begin{figure}[ht]
\setlength{\unitlength}{1mm}
\begin{picture}(150,115)
\put(-1,-2){\mbox{\epsfig{figure=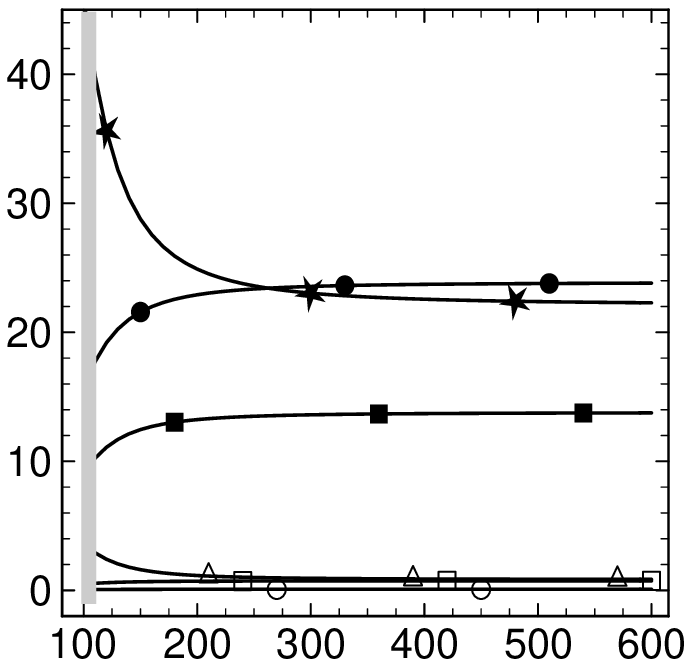,
                          height=11.2cm,width=13.5cm}}}
\put(0,109){\makebox(0,0)[bl]{{\large
             $\bf BR$~[\%]}}}
\put(133,-4){\makebox(0,0)[br]{{\large $\bf m_{A^0}$~[GeV]}}}
\end{picture}
\caption[]{Branching ratios of $\chin{2}$ as a function of $m_{A^0}$ 
for $\tan \beta = 4$ and $\mu = 500$~GeV.
The graphs correspond to following transitions:
$\circ \, \chin{2} \to \chin{1} \, \nu_l \, \nu_l$ ($l=e,\mu$, or $\tau$), 
$\square \, \chin{2} \to \chin{1} \, l^+ \, l^-$ ($l=e$ or $\mu$), 
$\triangle \, \chin{2} \to \chin{1} \, \tau^+ \, \tau^-$, 
$\blacksquare \, \chin{2} \to \chin{1} \, q \, \bar{q}$ ($q=u$ or $c$), 
$\bullet \,  \chin{2} \to \chin{1} \, q \, \bar{q}$ ($q=d$ or $s$), and 
$\star \,  \chin{2} \to \chin{1} \, b \, \bar{b}$.  In the gray area is
$m_{h^0} < 90$~GeV.}
\label{fig:BrMap}
\end{figure}

The $m_{A^0}$ dependence of the $\chin{2}$ decay branching ratios
is shown in \fig{fig:BrMap}. As can be seen the Higgs boson exchange is
important for $m_{A^0} \lsim 200$~GeV and decreases for larger $m_{A^0}$.
This is because $|\sin \alpha|$ decreases with increasing $m_{A^0}$
leading to a reduction the of $h^0 \, b \, \bar{b}$ coupling 
[Eq.~(\ref{coup:Hbb})].

\begin{figure}[ht]
\setlength{\unitlength}{1mm}
\begin{picture}(150,115)
\put(-1,-1){\mbox{\epsfig{figure=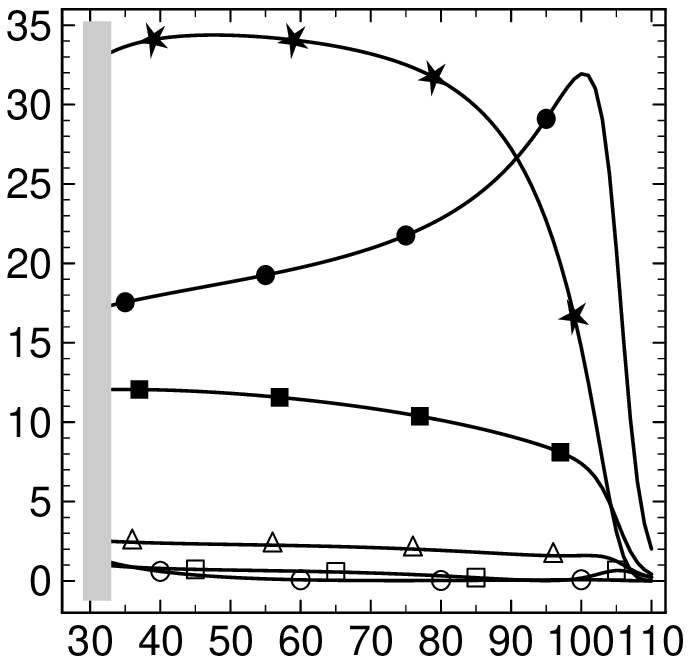,
                          height=11.0cm,width=13.5cm}}}
\put(0,109){\makebox(0,0)[bl]{{\large
             $\bf BR$~[\%]}}}
\put(134,-4){\makebox(0,0)[br]{{\large $\bf M'$~[GeV]}}}
\end{picture}
\caption[]{Branching ratios of $\chin{2}$ 
as a function of $M'$ (in GeV) for
$M_2 = 120$~GeV, $\mu = 500$~GeV, $\tan \beta =4$, and $m_{A^0} = 125$~GeV.
The other parameters are given in the text.
The graphs correspond to following transitions:
$\circ \, \chin{2} \to \chin{1} \, \nu_l \, \nu_l$ ($l=e,\mu$, or $\tau$), 
$\square \, \chin{2} \to \chin{1} \, l^+ \, l^-$ ($l=e$ or $\mu$), 
$\triangle \, \chin{2} \to \chin{1} \, \tau^+ \, \tau^-$, 
$\blacksquare \, \chin{2} \to \chin{1} \, q \, \bar{q}$ ($q=u$ or $c$), 
$\bullet \,  \chin{2} \to \chin{1} \, q \, \bar{q}$ ($q=d$ or $s$), and 
$\star \,  \chin{2} \to \chin{1} \, b \, \bar{b}$. In the gray area is
$m_{\chin{1}} < 30$~GeV.}
\label{fig:BrM1p}
\end{figure}

In \fig{fig:BrM1p} we show the dependence of the $\chin{2}$ branching
ratios on the $U(1)$ gaugino mass parameter $M'$ taking $M=500$~GeV,
$\mu=500$~GeV, and $\tanb=4$. This dependence is rather weak except
for $M' \sim M$ where $\mchin{2} \sim \mchip{1} \sim \mchin{1}$.

\section{Summary}

In this paper we have shown that the Higgs boson exchange contributions to
three-body decay modes of the second lightest neutralino can be very important
even if $\tan \beta$ is small.
This is valid for scenarios where the sfermions are heavy ($\gg m_Z$)
and $m_{A^0}$ is not too heavy ($m_{A^0} \lsim 200$~GeV). 
This affects the signatures for $\chin{2}$ decays at LEP, FNAL, LHC, 
an $e^+ e^-$ linear collider, and a muon collider. 
Most studies have been done within the minimal supergravity model
 \cite{sugra1,sugra2}. A feature of this model is that  $m_{A^0}$ is in
general large and consequently  $|\sin \alpha|$ relatively small, implying a 
small $h^0 - b \bar{b}$ coupling  for small $\tanb$. In a more general 
framework this is not necessarily true and the
 Higgs boson exchange contribution could be much larger.

\section*{Acknowledgments}

This work was supported by the ``Fonds zur F\"orderung der wissenschaftlichen
Forschung'' of Austria, project No.~P13139-PHY. W.P. was supported by
Spanish 'Ministerio de Educacion y Cultura' under the contract SB97-BU0475382.
M.~Kleander's help in checking some parts of the computer program is
gratefully acknowledged.


\newpage

\newpage

\newpage

\newpage

\newpage

\newpage

\end{document}